\documentclass[%
preprint,
superscriptaddress,
nofootinbib,
nobibnotes,
 amsmath,amssymb,
 prl,
]{revtex4-2}

\usepackage{graphicx}
\usepackage{dcolumn}
\usepackage{longtable}
\usepackage{bm}
\usepackage{comment}
\usepackage{lineno}
\usepackage[none]{hyphenat}
\usepackage{hyperref}
\hypersetup{hidelinks,colorlinks=true,citecolor=blue,linkcolor=blue,breaklinks=true,urlcolor=blue}
\usepackage{xcolor}
\usepackage{amsmath}

\usepackage{float}

\begin{document}

\title{
Superconducting Cavity-Based Sensing of Band Gaps in 2D Materials
}


\author{Krishnendu Maji}
\author{Joydip Sarkar}
\author{Supriya Mandal}
\author{Sriram H.}

\author{Mahesh Hingankar}
\author{Ayshi Mukherjee}
\author{Soumyajit Samal}
\author{Anirban Bhattacharjee}
\author{Meghan P. Patankar}
\affiliation{Department of Condensed Matter Physics and Materials Science, Tata Institute of Fundamental Research, Homi Bhabha Road, Mumbai 400005, India.}
\author{Kenji Watanabe}
\affiliation{Research Center for Functional Materials,
National Institute for Materials Science, 1-1 Namiki, Tsukuba 305-0044, Japan.}
\author{Takashi Taniguchi}
\affiliation{International Center for Materials Nanoarchitectonics, National Institute for Materials Science,  1-1 Namiki, Tsukuba 305-0044, Japan.}
\author{Mandar M. Deshmukh}
\email{deshmukh@tifr.res.in}
\affiliation{Department of Condensed Matter Physics and Materials Science, Tata Institute of Fundamental Research, Homi Bhabha Road, Mumbai 400005, India.}

\begin{abstract}
\textbf{The superconducting coplanar waveguide (SCPW) cavity plays an essential role in various areas like superconducting qubits, parametric amplifiers, radiation detectors, and studying magnon-photon and photon-phonon coupling. Despite its wide-ranging applications, the use of SCPW cavities to study various van der Waals 2D materials is relatively unexplored. The resonant modes of the SCPW cavity exquisitely sense the dielectric environment. In this work, we measure the charge compressibility of bilayer graphene coupled to a half-wavelength SCPW cavity. Our approach provides a means to detect subtle changes in the capacitance of the bilayer graphene heterostructure, which depends on the compressibility of bilayer graphene, manifesting as shifts in the resonant frequency of the cavity. This method holds promise for exploring a wide class of van der Waals 2D materials, including transition metal dichalcogenides (TMDs) and their moiré where DC transport measurement is challenging.}
\end{abstract}

\maketitle

\newpage

Superconducting coplanar waveguide (SCPW) cavities are structures designed to confine and manipulate electromagnetic fields at microwave frequencies\cite{goppl_coplanar_2008}, which are widely used in diverse areas like coupling to superconducting qubit for readout\cite{wallraff_strong_2004,wallraff_approaching_2005,schuster_resolving_2007,majer_coupling_2007}, parametric amplifiers\cite{tholen_nonlinearities_2007,castellanos-beltran_widely_2007}, bolometers\cite{day_broadband_2003,lee_graphene-based_2020}, fast charge sensing of quantum dots\cite{stehlik_fast_2015}, studying magnon-photon interaction in van der Waals magnets\cite{hou_strong_2019,mandal_coplanar_2020}, studying interaction with spin ensemble\cite{amsuss_cavity_2011,kubo_strong_2010,ranjan_probing_2013}, and studying nanomechanical devices\cite{regal_measuring_2008,singh_optomechanical_2014,sahu_superconducting_2022}. The SCPW cavities when operated at resonant frequency are very sensitive and find application in detecting small changes in the electromagnetic field\cite{schmidt_current_2020,endo_ghz_2014}, and radiation\cite{de_visser_fluctuations_2014}. Consequently, SCPW cavities offer powerful and sensitive tools for investigating the electronic properties of van der Waals materials.


van der Waals 2D materials consist of atomically thin layers stacked on top of each other through weak van der Waals force. These materials exhibit extraordinary electronic, optical, and mechanical properties due to their reduced dimensionality. Furthermore, the stacking of various van der Waals materials at specific angles between layers gives rise to new functionalities and electronic properties\cite{cao_correlated_2018,cao_unconventional_2018,ma_moire_2020,pan_quantum-confined_2018}. While various measurements have been conducted to study these materials, most have focused on low-frequency DC transport measurements. One of the key limitations of DC transport measurement is that it requires ohmic contact between the electrodes and 2D materials. There is a wide range of 2D materials in which establishing ohmic contact is challenging. The significant advantage of the microwave technique is that it enables contactless transport measurements, such as measuring the electrical conductivity of 2D heterostructure through the transmission of a capacitively coupled cavity, without the need for physical contacts or electrodes that might disrupt the behavior of 2D materials. 
The microwave cavity-based techniques exploit the extreme sensitivity of the resonant frequency to the effective dielectric or magnetic environment, and the high quality factor of the cavity improves this sensitivity. Numerous studies have employed microwave techniques to analyze low-dimensional materials. These studies encompass diverse applications, including measuring the charge occupation of quantum dots\cite{schoelkopf_radio-frequency_1998}, assessing minute capacitance\cite{malinowski_radio-frequency_2022}, and measuring the impedance of low-dimensional materials\cite{johmen_radio-frequency_2023}, exploring the microwave properties of graphene-based Josephson junction and their potential applications\cite{schmidt_ballistic_2018,dou_microwave_2021,haller_phase-dependent_2022,kroll_magnetic_2018,lee_graphene-based_2020,kokkoniemi_bolometer_2020,sarkar_quantum-noise-limited_2022,butseraen_gate-tunable_2022,wang_coherent_2019}.

In our work, we show an effective application of SCPW cavities to explore electronic compressibility, a parameter related to the density of states (DOS) in bilayer graphene (BLG). The BLG heterostructure is used as one of the coupling capacitors of a half-wave ($\lambda/2$) cavity with the ability to apply gate voltages to tune its band structure\cite{zhang_direct_2009}. Specifically, we extract the capacitance of the BLG stack from the transmitted signal through the cavity, where the resonant frequency shifts with the capacitance of the stack. This capacitance measurement allows us to extract the compressibility of the BLG. Moreover, the dual gate geometry of the device allows independent control of the charge density ($n$) in the BLG and perpendicular displacement field ($D$). The displacement field tunes the band gap of the BLG, thus modifying its DOS. Our compressibility measurement can capture this small opening of the band gap. Our method could be useful in studying various other 2D materials, and bilayer graphene serves as a demonstrator for the study of materials with small tunable gaps in the electronic spectrum. Extending our work to other 2D heterostructures could offer a highly sensitive and non-intrusive means of investigation, further enhancing our understanding of various 2D materials' properties. Most of the existing techniques to measure the capacitance of low-dimensional materials rely on lumped element circuits operating within a few hundred kHz range\cite{young_electronic_2012,henriksen_measurement_2010}. Whereas, our SCPW cavity operates at a few GHz frequency. The choice of GHz frequency range allows scopes for fast dynamical measurements in a cavity-coupled system\cite{amsuss_cavity_2011,stehlik_fast_2015,pirkkalainen_hybrid_2013} and can be extended to 2D material platforms as well\cite{yoo_time_2023}. A theoretical study has recently been conducted on quantum geometry-originated capacitance in insulators\cite{komissarov_quantum_2023}; measuring the capacitance of flat-band moiré systems could provide insights into the quantum geometry. In addition, the moiré platform has been proposed as a route toward the quantum simulation of correlated electronic states \cite{kennes_moire_2021}. Our technique can be useful to probe many of these quantum simulators experimentally.

\begin{figure}
    \centering
    \includegraphics[width=0.95\textwidth]{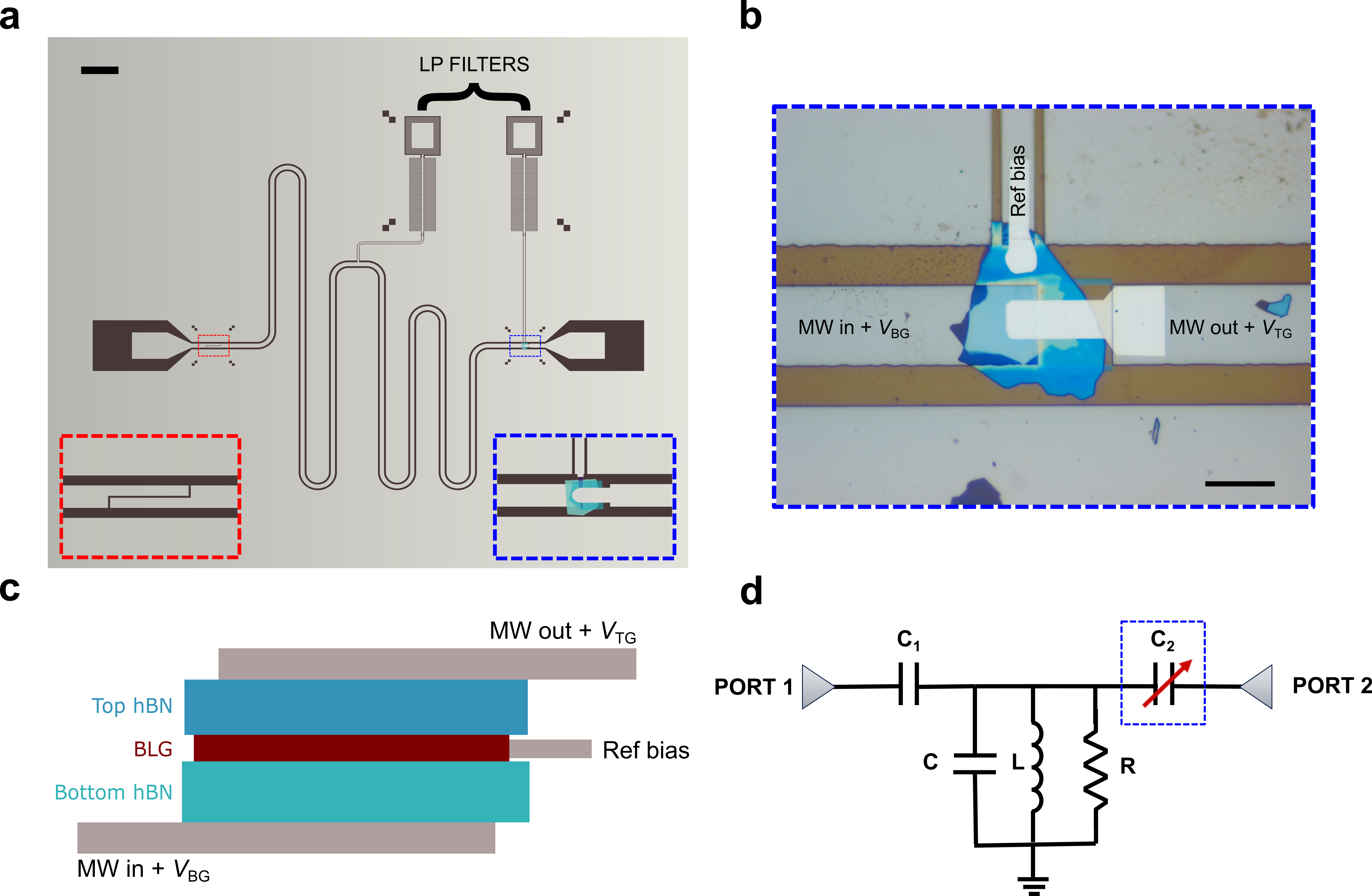}
    \caption{ \textbf{Coupling the 2D heterostructure to a superconducting cavity.} 
    \textbf{a,} The representative schematic of the device. The $\lambda$/2 cavity is designed on a MoRe-sputtered  SiO$_{\mathrm{2}}$ (280 nm)/intrinsic Si (500 $\mathrm{\mu m}$) substrate. The inset at the bottom left shows the finger coupling capacitor designed on the left side of the cavity. The inset on the bottom right side shows the zoomed-in image of the BLG heterostructure which serves as the second coupling capacitor $\mathrm{C_2}$. The scale bar on the top left is 200 $\mathrm{\mu m}$. 
    \textbf{b,} The optical image of the BLG heterostructure. The microwave (MW) signal and applied DC voltage schemes are indicated in the figure. The scale bar is 20 $\mathrm{\mu m}$.
    \textbf{c,} The cross-sectional schematic of the BLG heterostructure. 
    \textbf{d,} Schematic diagram of the equivalent circuit. The BLG stack plays the role of coupling capacitor $\mathrm{C_2}$ marked by the blue dashed box.
    }
    \label{fig:fig1} 
\end{figure}

Figure \ref{fig:fig1} a shows a schematic representation of the fabricated device. The device is fabricated on SiO$_{\mathrm{2}}$ (280 nm)/intrinsic Si (500 $\mathrm{\mu m}$) substrate. The substrate is initially DC magnetron sputtered with 50 nm MoRe, a type-$\mbox{II}$ superconductor. Following this, photolithography and reactive ion etching are used to pattern the $\lambda$/2 SCPW cavity on the substrate. The cavity is designed to have a resonant frequency of around 6 GHz and a characteristic impedance of 50 $\Omega$. The cavity is coupled to the input and output ports via coupling capacitors for transmission measurements. A finger coupling capacitor of 3.4 fF is designed at the input side\cite{goppl_coplanar_2008}. At the output side, the van der Waals heterostructure serves as the coupling capacitor. In our work, the stack consists of a BLG flake encapsulated between a top hBN and a bottom hBN (see Supplementary Information, section S1 for details). Figure \ref{fig:fig1} b shows the optical image of the heterostructure. The cross-section of the heterostructure is schematically shown in Figure \ref{fig:fig1} c. The bottom gate voltage is applied through a low-pass (LP) filter linked to the CPW's central conductor at the voltage node to minimize microwave loss\cite{mi_circuit_2017}. A bias tee is used to apply the top gate voltage through the output microwave port (see Supplementary Information, section S2 for measurement details). The BLG is biased to the zero potential through another LP filter. The ability to apply multi-terminal DC voltages is an added advantage as many moiré heterostructures have a band structure that is widely tunable with displacement field. The equivalent circuit of the device is illustrated in Figure \ref{fig:fig1} d, where the heterostructure plays the role of the tunable coupling capacitor $\mathrm{C_2}$, which is marked by the blue dashed box. 

\begin{figure}
    \centering
    \includegraphics[width=0.6\textwidth]{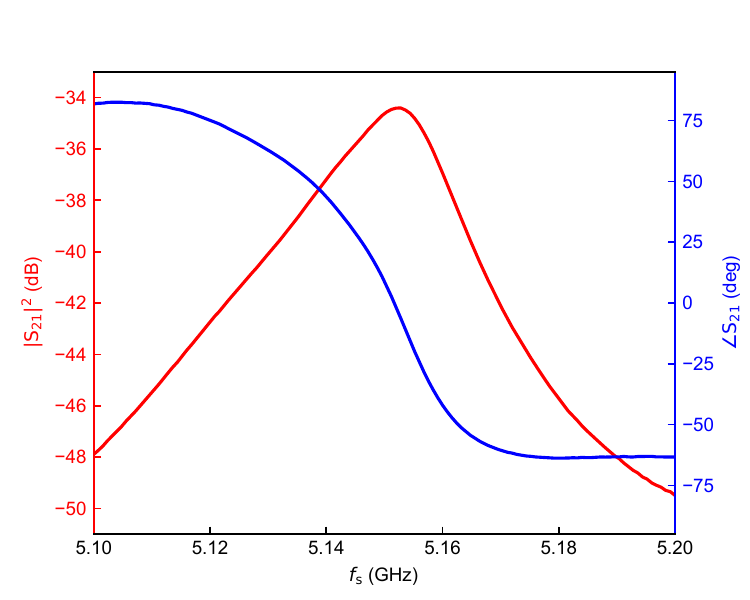}
    \caption {\textbf{Cavity characterization.} The red curve and the blue curve show the amplitude ($|\mathrm{S}_{21}|^2$) and phase ($\angle\mathrm{S}_{21}$) of the transmission coefficient respectively of the cavity at zero gate voltages. The resonant frequency is around 5.152 GHz. The cavity is showing a $\pi$ phase change at the resonance, which is the characteristic of a $\lambda /2$ cavity measured in transmission mode.}
    \label{fig:fig2}
\end{figure}

The measurements are done in a dilution fridge at 20 mK. The device is loaded inside an aluminum puck to eliminate stray magnetic fields and create a better electromagnetic environment at the base temperature. The transmission coefficient S$_\mathrm{21}$, the ratio of the transmitted signal to the input signal, is measured as a function of frequency using a vector network analyzer (VNA). The amplitude and phase of the transmission coefficient, S$_\mathrm{21}$, is plotted in Figure  \ref{fig:fig2} as a function of signal frequency at zero gate voltages. The $\pi$ phase change is the characteristic of a $\lambda /2$ cavity measured in transmission mode. The resonant frequency of the cavity is around 5.152 GHz. In Supplementary Information, Figure S2, the COMSOL simulation result of the cavity is shown.

\begin{figure}
    \centering
    \includegraphics[width=0.95\textwidth]{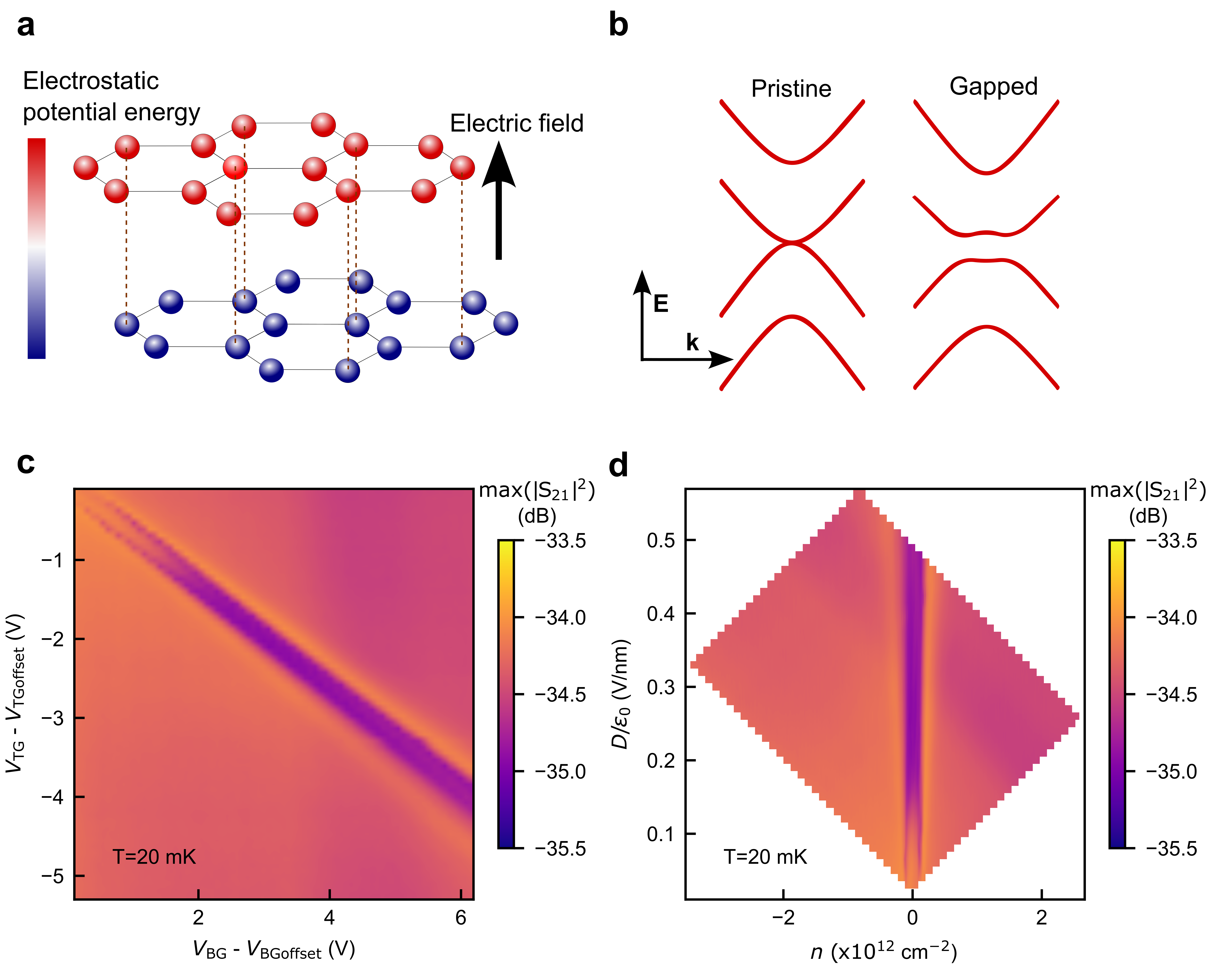}
    
    \caption{\textbf{The transmitted signal through the cavity as a function of the gate voltages.}
    \textbf{a,} Upon applying a perpendicular displacement field, an electrostatic potential energy difference is created between the two layers of bilayer graphene, resulting in the breaking of inversion symmetry.
    \textbf{b,} Left: The schematic shows that the pristine bilayer graphene has a zero band gap in its electronic structure. Right: Upon application of a perpendicular displacement field, a gap opens up due to the breaking of inversion symmetry.
    \textbf{c,}  For each ($V_{\mathrm{TG}}$, $V_{\mathrm{BG}}$), the transmitted signal through the cavity is measured as a function of frequency. The peak value of the amplitude of the transmission coefficient, 
   max($|\mathrm{S}_{21}|^2$) is plotted as a function of  $V_{\mathrm{BG}}$ and $V_{\mathrm{TG}}$. There is a finite offset in both the $V_{\mathrm{BG}}$ and $V_{\mathrm{TG}}$ axes due to charge doping in BLG from the metallic gates. The measurements are done at 20 mK. \textbf{d,} max($|\mathrm{S}_{21}|^2$) is plotted as a function of $n$ and $D/\epsilon_0$. The region of low transmission along the displacement field corresponds to the opening of the band gap in BLG.}
   \label{fig:fig3}
\end{figure}
 The device is fabricated in a dual-gate configuration, which enables independent control over the carrier density ($n$) of the BLG and the perpendicular displacement field ($D$) by adjusting both the top gate voltage ($V_{\mathrm{TG}}$) and the bottom gate voltage ($V_{\mathrm{BG}}$). In addition, the bilayer graphene is biased to 0~V to provide a ground reference for the DC electrostatic tuning of the system. In pristine BLG, the two graphene layers are electrically identical, making it an inversion-symmetric system. Consequently, there is no band gap in its electronic structure as shown in Figure \ref{fig:fig3} b. Upon applying a perpendicular displacement field, an electrostatic potential difference is created between the two layers (shown in Figure \ref{fig:fig3} a), breaking the inversion symmetry and resulting in the opening of a band gap (shown in Figure \ref{fig:fig3} b).
 
 There are two ways to probe the system experimentally. First, one measures the $\mathrm{S}_{21}$ frequency response of the cavity-coupled device at each of the values of $V_{\mathrm{TG}}$ and $V_{\mathrm{BG}}$ creating a 4-dimensional dataset, and second, by observing cavity response at a frequency close to the resonant mode of the system and varying the gate voltages creating a 3-dimensional dataset. In this work, we focus on the first method of probing the system. For each ($V_{\mathrm{TG}}$, $V_{\mathrm{BG}}$), $\mathrm{S}_\mathrm{21}$ as a function of frequency is measured. Then the peak value of $|\mathrm{S}_{21}|^2$ for each ($V_{\mathrm{TG}}$, $V_{\mathrm{BG}}$) is extracted from the data, which is plotted in Figure \ref{fig:fig3} c. The data is also taken in the second way where the transmission signal is measured at a fixed frequency close to the resonance (see Supplementary Information, Figure S3). In our data, an offset of $V_{\mathrm{BGoffset}}$ = -1.15 V and $V_{\mathrm{TGoffset}}$ = -2.7 V is observed in  $V_{\mathrm{BG}}$ and $V_{\mathrm{TG}}$, respectively, likely due to the charge doping in graphene from the metallic top and bottom gate which has been previously reported\cite{nouchi_determination_2011}. The data offer clearer physical insights if plotted in ($n$, $D$) space instead of ($V_{\mathrm{BG}}$, $V_{\mathrm{TG}}$). The transformation from ($V_{\mathrm{BG}}$, $V_{\mathrm{TG}}$) to ($n$, $D$) is achieved using the following set of equations: $n = (C_{\mathrm{BG}}V_{\mathrm{BG}}' + C_{\mathrm{TG}}V_{\mathrm{TG}}')/e$ and $D = (C_{\mathrm{BG}}V_{\mathrm{BG}}' - C_{\mathrm{TG}}V_{\mathrm{TG}}')/2$, where $C_{\mathrm{BG}}$ ($C_{\mathrm{TG}}$) is the capacitance of the bottom (top) gate, $V_{\mathrm{BG}}' (V_{\mathrm{TG}}') = V_{\mathrm{BG}} (V_{\mathrm{TG}}) - V_{\mathrm{BGoffset}} (V_{\mathrm{TGoffset}})$  is the offset corrected bottom (top) gate voltage, and $e$ is the electronic charge. The plot in Figure \ref{fig:fig3} c is transformed to ($n$, $D/\epsilon_0$) plane from ($V_{\mathrm{BG}}$,  $V_{\mathrm{TG}}$) plane in Figure \ref{fig:fig3} d. Only the positive displacement field sector data is shown here (see Supplementary Information, Figure S4 for the full range data). A region with low transmission appears along the electric displacement field axis. This region corresponds to the opening of a band gap in the BLG caused by the perpendicular displacement field\cite{zhang_direct_2009}. The width of the low transmission region at a particular displacement field relates to the band gap of BLG at that displacement field. On looking closely at the data, one can see in the low transmission region two subfeatures emerge. These subfeatures correspond to two charge neutrality points and can be attributed to the spatial inhomogeneous doping in the device, a consequence of its larger overall area of $\sim$30 $\mathrm{\mu}$m$^2$. In the data shown in Figure \ref{fig:fig3} c and d, a van-Hove singularity-like feature appears at the band edge of the BLG, which is a characteristic of BLG band structure (see Supplementary Information, section S4.3 for details).

To extract the capacitance of the heterostructure, the data is fitted with the scattering coefficients $\mathrm{S}_{21}$ of a capacitively coupled $\lambda /2$ CPW cavity, which can be written as\cite{chen_scattering_2022}
\begin{equation}
    \mathrm{S}_{21} = \frac{2}{A + B/Z_0 + CZ_0 + D} + b,
    \label{Eq:eq1}
\end{equation}
where $A$, $B$, $C$, and $D$ are the elements of the transmission matrix or ABCD matrix \newline 
$A = \cosh(\gamma l)$ + $\frac{\sinh(\gamma l)}{j\omega C_1 Z_0}$,~~~~~$B = \sinh(\gamma l)\left(Z_0-\frac{1}{\omega^2 C_1 C_2 Z_0}\right) + \cosh(\gamma l)\left(\frac{1}{j\omega C_1}+\frac{1}{j\omega C_2}\right)$ \newline
$C = \frac{\sinh(\gamma l)}{Z_0}$,~~~~~$D = \cosh(\gamma l) + \frac{\sinh(\gamma l)}{j\omega C_2 Z
_0}$
\newline
where $l$ is the length of the transmission line, $\gamma = \alpha + j\beta$ is the complex propagation constant of the microwave field, $Z_0$ is the characteristic impedance of the transmission line. For $\lambda/2$ cavity $\beta l = \pi + \pi (\omega-\omega_0)/\omega_0$, where $\omega_0$ is the bare resonant frequency and $\alpha l = \pi/(2Q_i)$, where $Q_i$ is the internal quality factor of the cavity. $b$ accounts for and adjusts the reference value of the transmission signal.

\begin{figure}
    \centering
    \includegraphics[width=0.95\textwidth]{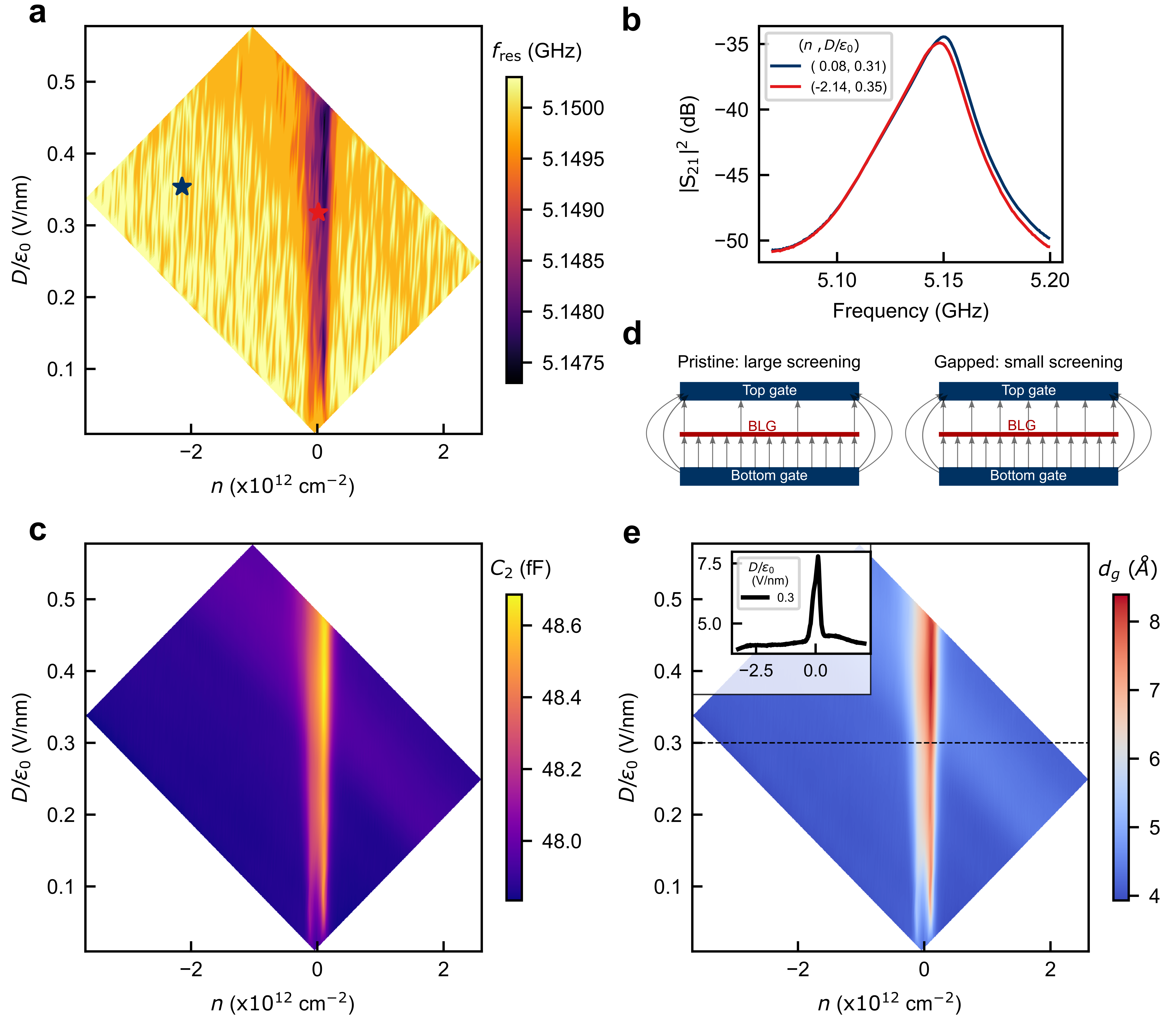}
    \caption{ \textbf{Measurement of the capacitance of the BLG heterostructure.} \textbf{a,} Shows the change in resonant frequency as a function of $n$ and $D/\epsilon_0$. The resonant frequency is shifting downward with the gap opening in BLG. \textbf{b,} Shows the frequency line-slice at two ($n$, $D/\epsilon_0$) points as marked in panel \textbf{a}.
    \textbf{c,} Shows the extracted capacitance $C_2$ from the fitting as a function of $n$ and $D/\epsilon_0$. The capacitance increases with gap opening in the BLG, which is a consequence of the decreased density of states (DOS) of the BLG, which can also be interpreted from eq \ref{Eq:eq2}.
    \textbf{d,} Left: In pristine BLG when there is no band gap in its electronic structure, the DOS of BLG is high. The BLG acts more like a metal in this regime, thus effectively screening the electric field between the top and bottom gates. Right: With a gap opening in BLG induced by a perpendicular displacement field, the DOS of BLG significantly decreases, transforming BLG into more of an insulator. Consequently, the screening effect is less prominent.
    \textbf{e,} Shows the extracted $d_g$ from the panel \textbf{c} using eq \ref{Eq:eq2}, where $d_g$ is related to the inverse DOS ($\frac{1}{e^2}\frac{\partial\mu}{\partial n}$) of BLG as  $d_g=\frac{\epsilon_0}{e^2}\frac{\partial\mu}{\partial n}$. The inset shows the line slice of $d_g$ with $n$ at $D/\epsilon_0$=0.3 V/nm.
    }
    \label{fig:fig4} 
\end{figure}
In Figure \ref{fig:fig4} a, the resonant frequency, \textit{i.e.}, the frequency at which the amplitude of $S_{21}$ is maximum is plotted in the ($n$, $D/\epsilon_0$) plane. In Figure \ref{fig:fig4} b, the frequency line-slice for two ($n$, $D/\epsilon_0$) points are shown. When the BLG has an opening in the gap, its resonant frequency shifts downward, and this frequency shift becomes more pronounced as the band gap increases.


For each ($n$, $D/\epsilon_0$), the $\mathrm{S}_{21}$ frequency response is fitted using eq \ref{Eq:eq1}. To address the observed asymmetry in the frequency line-slice data caused by background effects, a skewness term $\left(1+erf \left(s(f-f_0)\right)\right)$ is multiplied with the fitting function, where $erf$ is the error function and $s$ is the skewness factor.
The extracted capacitance value $C_2$ from the fitting is plotted in Figure \ref{fig:fig4} c (for details of the fitting procedure see Supplementary Information, section S5). This capacitance value, which increases when a gap opens in the BLG, directly correlates with the DOS of BLG. Using our cavity-based technique, subfemtofarad change in capacitance can be measured accurately (see Supplementary Information, Figure S6). The accuracy of our capacitance measurement has been compared with other state-of-the-art techniques\cite{young_electronic_2012,henriksen_measurement_2010} (see Supplementary Information, section S7 for details). 

The extracted capacitance $C_2$ represents the penetration field capacitance between the top gate and bottom gate, where grounded BLG screens the electric field due to its finite DOS\cite{henriksen_measurement_2010,young_capacitance_2011}. By considering the BLG as an infinitely thin sheet, the expression for the measured capacitance can be written as\cite{henriksen_measurement_2010}
\begin{equation}
\label{Eq:eq2}
    C_2=\frac{\epsilon_0 A_t d_g \epsilon_{\mathrm{hBN}}^2}{d_g \epsilon_{\mathrm{hBN}}\left(d_b + d_t\right)+\epsilon_0 d_b d_t} + C_S.
\end{equation}
Here $A_t$ is the area of the top gate covering the BLG; $d_b$ and $d_t$ are the thickness of the bottom hBN and top hBN, respectively; $\epsilon_{\mathrm{hBN}}$ and $\epsilon_0$ are the permittivity of hBN and free space, respectively. $d_g=\frac{\epsilon_0}{e^2}\frac{\partial\mu}{\partial n}$ is the parameterization of $\frac{1}{e^2}\frac{\partial\mu}{\partial n}$ in units of length, where $\frac{\partial\mu}{\partial n}$ is the inverse DOS of the BLG. The inverse DOS is related to the electronic compressibility (K) as $K^{-1}=n^2\frac{\partial\mu}{\partial n}$. The stray capacitance $C_S$, assumed to be 47.81 fF, is estimated using the area of the top gate 110 $\mu\mathrm{m}^2$ and the $d_t= 32.9 \pm  0.6$ nm and $d_b=50.1 \pm 1$ nm. These thickness values have been measured using atomic force microscopy (AFM; see Supplementary Information, section S6 for details). From eq \ref{Eq:eq2} one can see that the $C_2$ increases as the DOS decreases. In the pristine BLG, the DOS is high when there is no band gap in its electronic structure. As a result, BLG screens the electric field between the top and bottom gates more effectively resulting in a decrease in the $C_2$ (as depicted in Figure \ref{fig:fig4} d). Upon application of a perpendicular displacement field, a gap opens in BLG, substantially reducing the DOS. Consequently, the screening is less effective for gapped BLG (as shown in Figure \ref{fig:fig4} b) and $C_2$ increases.

From the extracted capacitance $C_2$, $d_g$ is obtained using eq \ref{Eq:eq2} and is shown in Figure \ref{fig:fig4} e. The value of $d_g$ matches well with previous reports\cite{henriksen_measurement_2010}. 
The line slice of the extracted $d_g$ at $D/\epsilon_0$=0.3 V/nm is shown in the inset of Figure \ref{fig:fig4} e. 
The $d_g$ has a sharp peak at $n=0$ and decreases asymmetrically on either side of $n$ =0. The peak in $d_g$ can be interpreted as the emergence of a band gap in bilayer graphene (BLG), resulting in a reduced DOS. The data can be explained by a minimal tight-binding model of BLG. The inverse DOS has been calculated by solving the tight-binding Hamiltonian with a gap size $\Delta$, which agrees well with the experimental data (see Supplementary Information, section S8 for details)\cite{mccann_electronic_2013}.

In summary, in this study, a half-wave SCPW cavity is used to characterize the electronic compressibility of BLG, establishing a direct correlation between the transmitted signal through the cavity and the band gap in BLG as a function of the displacement field. Also, from the shift in resonant frequency, the capacitance of the stack is extracted, which directly correlates with the inverse DOS of the BLG. This noncontact microwave cavity-based technique provides a scope for noninvasive sensing of DOS in 2D materials.

Our methodology can be easily extended to the study of other 2D materials, as well. Furthermore, this technique holds promise for gaining insights into moiré superlattices of transition metal dichalcogenides, which have gained significant attention as a promising platform as a quantum simulator\cite{kennes_moire_2021}. Importantly, our technique overcomes the challenge of contact resistance, which hampers the study of many exciting 2D materials. In addition, the choice of microwave frequencies will also enable fast dynamical measurements\cite{amsuss_cavity_2011,stehlik_fast_2015,pirkkalainen_hybrid_2013} in a cavity-coupled 2D van der Waals system.

\section{Acknowledgements}

We thank R. Vijay and V. Singh for their helpful discussions and 
comments. We thank H. Agarwal and D. Jangade for helping with device fabrication and S. Layek for helping with data analysis. We thank U. Ghorai for helping with the tight binding calculation. We acknowledge 
Nanomission grant SR/NM/NS-45/2016 and the DST SUPRA 
SPR/2019/001247 grant, along with the Department of Atomic 
Energy of Government of India (12-R$\&$D-TFR-5.10-0100) for support. 
We also acknowledge support from the Department of Science and 
Technology, India, via the QuEST Programme. MMD acknowledges support from J.C. Bose Fellowship JCB/2022/000045 from the Department of Science and Technology of India. Preparation of hBN 
single crystals were supported by the Elemental Strategy Initiative 
conducted by the Ministry of Education, Culture, Sports, Science 
and Technology, Japan (grant number JPMXP0112101001) and Japan 
Society for the Promotion of Science KAKENHI (grant numbers 
19H05790 and JP20H00354).

\bibliography{bibliography}

\renewcommand{\thefigure}{S\arabic{figure}}

\setcounter{figure}{0}

\newpage
\begin{center}
   \section{Supplementary Information: Superconducting cavity-based sensing of band gaps in 2D materials} 
\end{center}

\section{\label{sec:level1}S1. Device fabrication}
\subsection{S1.1. Fabrication of the cavity}
The device is fabricated on SiO$_{\mathrm{2}}$ (280 nm)/intrinsic Si (500 $\mathrm{\mu m}$) substrate. The substrate is initially DC magnetron sputtered with 50 nm MoRe, a type-$\mbox{II}$ superconductor in a high-vacuum chamber ($\sim 3 \times 10^{-7}$ mbar) with a sputtering pressure of $\sim 2\times 10^{-3}$ mbar. Then, photolithography and SF$_6$ (12.5 sccm)/Ar (10 sccm) reactive ion etching are used to pattern the $\lambda$/2 SCPW cavity onto the substrate. For the fabrication of the LP filters electron beam lithography is used instead of photolithography.
\subsection{S1.2. Fabrication of the BLG heterostructure}
To make the hBN-BLG-hBN stack we exfoliate graphene and hBN flakes using scotch tape mechanical exfoliation technique. We choose suitable bilayer graphene flakes based on optical contrast. The hBN flakes of suitable thicknesses are selected based on color under an optical microscope and later thickness is confirmed by AFM after the device is completed. Subsequently, the flakes are assembled individually using poly(bisphenol A carbonate)/polydimethylsiloxane stamps. Then we drop the stack on the cavity. Next, we use standard electron-beam lithography followed by Aluminum deposition by evaporation at high vacuum ($\sim 2.5 \times 10^{-7}$) to make the top gate and electrode to bias the BLG.

\section{\label{sec:level2}S2. Measurement details}

\begin{figure}
    \centering
    \includegraphics[width=0.9\linewidth]{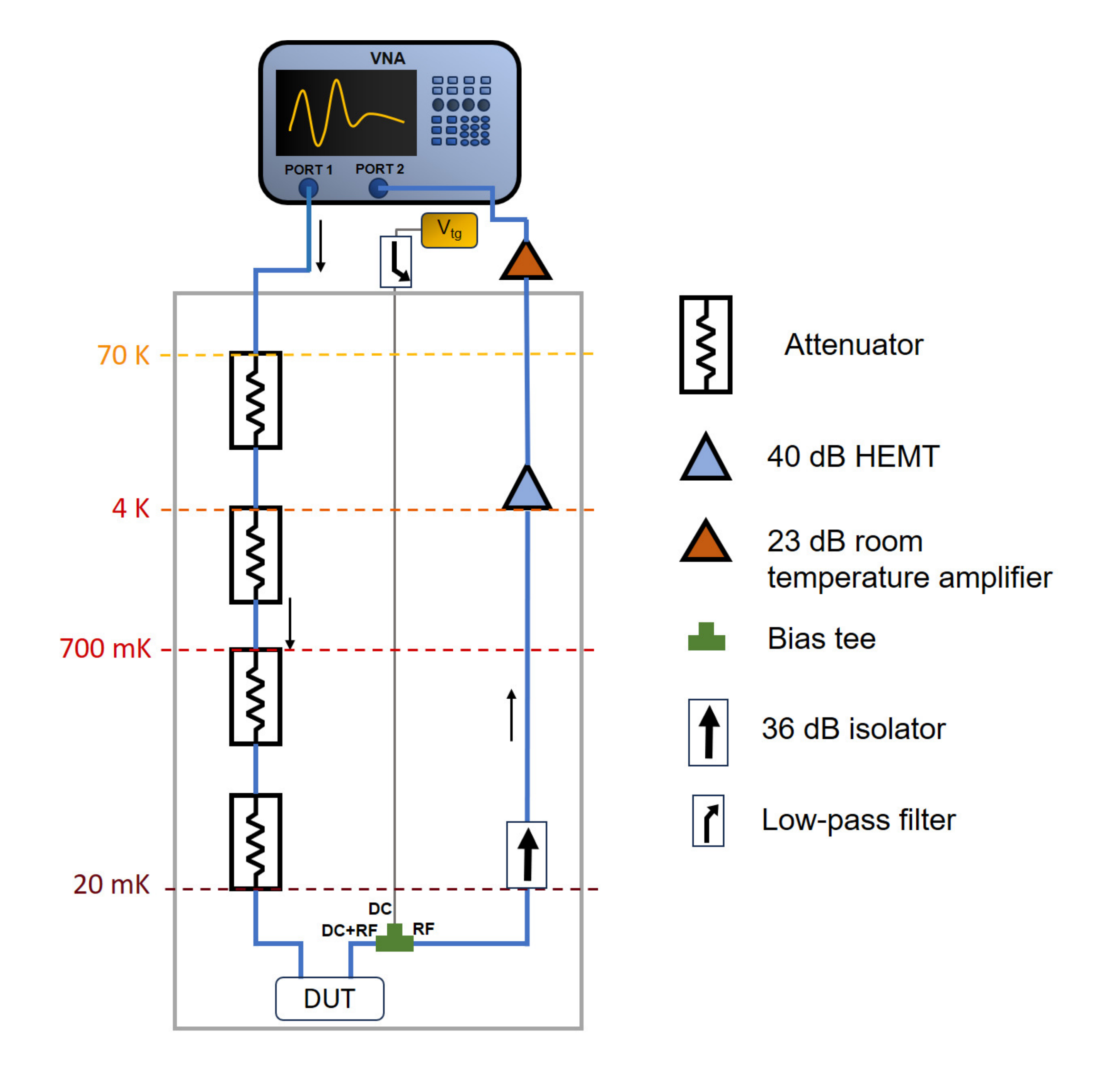}
    \caption {\textbf{Microwave measurement setup.} Shows the wiring diagram for the measurement. The input microwave signal is sent to the device from port 1 of the VNA through a series of attenuators kept at different plates of the dilution fridge. The output signal is amplified through a 40 dB high-electron-mobility-transistor (HEMT) amplifier followed by a 23 dB room temperature amplifier before reaching port 2 of the VNA. The DC voltage is applied to the top gate through a 10 Hz low-pass filter.} 
    \label{fig:Sup fig 1}
\end{figure}

Figure \ref{fig:Sup fig 1} show the circuit diagram of the measurement. The measurements are done in an Oxford dilution fridge at 20 mK. The input signal is attenuated by 60 dB using a series of attenuators kept at different plates of the dilution fridge for proper thermalization of the photons reaching the device. The output signal is amplified through a 40 dB high-electron-mobility-transistor (HEMT) amplifier kept at the 4 K plate, followed by a 23 dB room temperature amplifier (Mini-circuits). The measurements are done using a Rohde and Schwarz vector network analyzer (ZNB20). The DC voltage to the top gate is applied through a 10 Hz low-pass filter and combined with the output RF line using a bias tee. The DC voltage to the back gate is applied similarly which is not shown in the schematic.

\section{\label{sec:level3}S3. COMSOL simulation}
\begin{figure}
    \centering
    \includegraphics[width=0.9\linewidth]{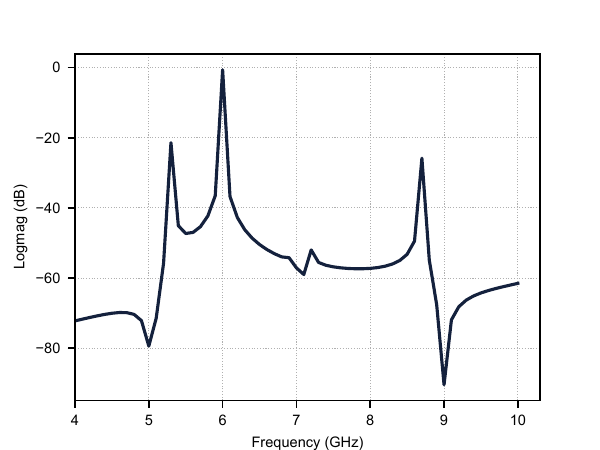}
    \caption {\textbf{COMSOL simuation} Simulated amplitude of the transmission coefficient ($|\mathrm{S}_{21}|^2$) of the cavity. The simulation was done in COMSOL Multiphysics software.}.   
    \label{fig:Sup fig 2}
\end{figure}

The finite element simulation has been performed using COMSOL to find the cavity modes. In Figure \ref{fig:Sup fig 2} the amplitude of the simulated transmission coefficient, $|\mathrm{S}_{21}|^2$ of the cavity is shown. The transmission peak which touches 0 dB is the mode of our interest, and it matches well with the experimental data.

\section{\label{sec:level4}S4. Additional data}

\subsection{S4.1. Data at a fixed frequency}
\begin{figure}
    \centering
    \includegraphics[width=\linewidth]{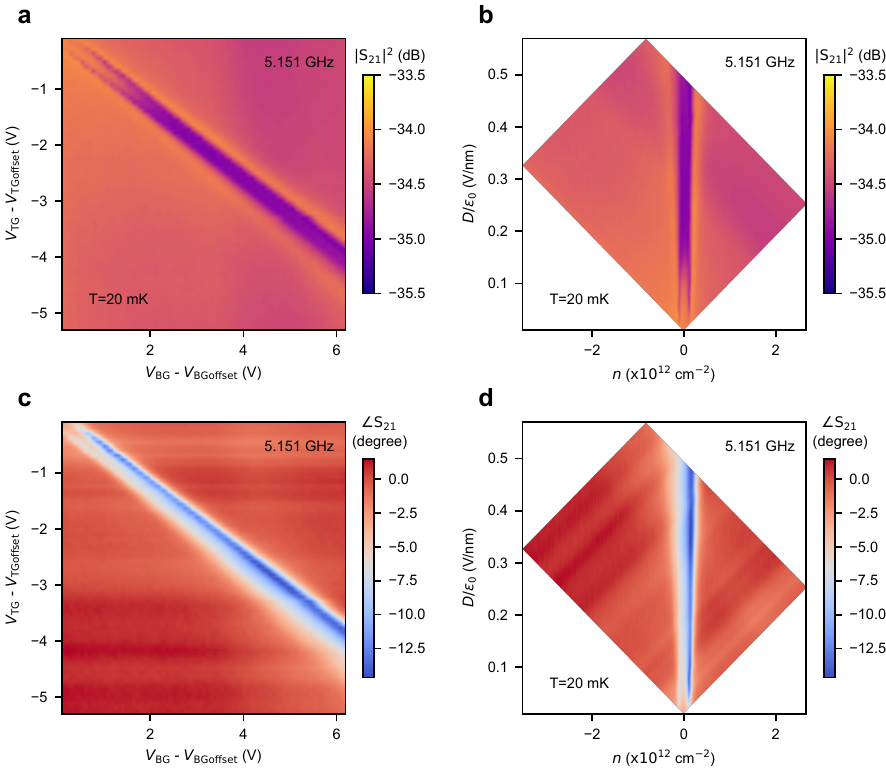}
    \caption {\textbf{Additional data.} 
    \textbf{a,}The amplitude of the transmission coefficient, $|\mathrm{S}_{21}|^2$ at frequency 5.151 GHz is plotted as function of ($V_{\mathrm{TG}}$, $V_{\mathrm{BG}}$). There is a finite offset in both $V_{\mathrm{BG}}$ and $V_{\mathrm{TG}}$ axes due to charge doping in BLG from the metallic gates.
    \textbf{b,} The data in panel \textbf{a} is shown in ($n,~ D/\epsilon_0$) plane. 
    \textbf{c,} The phase of the transmission coefficient, $\angle\mathrm{S}_{21}$ at frequency 5.151 GHz is plotted as a function of ($V_{\mathrm{TG}}$, $V_{\mathrm{BG}}$). 
    \textbf{d,} The data in panel \textbf{c} is shown in ($n,~ D/\epsilon_0$) plane.}   
    \label{fig:Sup fig 3}
\end{figure}

For each ($V_{\mathrm{TG}}$, $V_{\mathrm{BG}}$), transmission is measured at a frequency of 5.151 GHz. In Figure \ref{fig:Sup fig 3} a the amplitude of the transmission coefficient, $|\mathrm{S}_{21}|^2$ is plotted as a function of ($V_{\mathrm{TG}}$, $V_{\mathrm{BG}}$). Notably, there is a finite offset in both $V_{\mathrm{BG}}$ and $V_{\mathrm{TG}}$ axes due to charge doping from the metallic gates which have been discussed in detail in the main text. The data in Figure \ref{fig:Sup fig 3} a is transformed to ($n$, $D/\epsilon_0$) plane in Figure \ref{fig:Sup fig 3} b. In Figure \ref{fig:Sup fig 3} c the phase of the transmission coefficient, $\angle\mathrm{S}_{21}$ is plotted as a function of ($V_{\mathrm{TG}}$, $V_{\mathrm{BG}}$). The data in Figure \ref{fig:Sup fig 3} c is transformed to ($n$, $D/\epsilon_0$) plane in Figure \ref{fig:Sup fig 3} d. 

\subsection{S4.2. Data in the negative displacement field sector}
\begin{figure}[H]
    \centering
    \includegraphics[width=\linewidth]{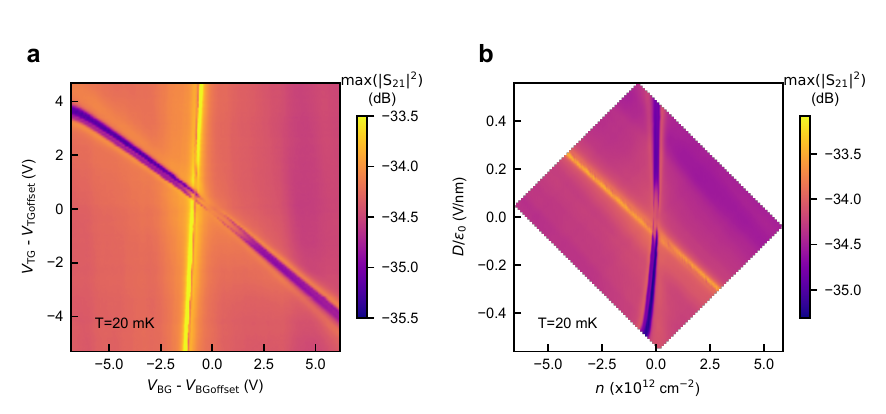}
    \caption {\textbf{Data in both the positive and negative displacement field sectors.}
    \textbf{a,} For each ($V_\mathrm{TG}$, $V_{\mathrm{BG}}$) the transmitted signal through the cavity is measured as a function of frequency. The peak value of the amplitude of the transmission coefficient, $|\mathrm{S}_{21}|^2$ is plotted as a function of $V_{\mathrm{TG}}$ and $V_{\mathrm{BG}}$. A finite offset in both $V_{\mathrm{TG}}$ and $V_{\mathrm{BG}}$ axes is due to charge doping in BLG from the metallic gates. \textbf{b,} max($|\mathrm{S}_{21}|^2$) is plotted as a function of $n$ and $D/\epsilon_0$.}

    \label{fig:Sup fig 4}
\end{figure}

The data was collected in the negative displacement field sector also. In Figure \ref{fig:Sup fig 4} a, we have plotted the full data in the ($V_{\mathrm{BG}}$, $V_{\mathrm{TG}}$) plane. The plot in Figure \ref{fig:Sup fig 3} a is transformed to ($n$, $D/\epsilon_0$) plane from ($V_{\mathrm{BG}}$,  $V_{\mathrm{TG}}$) plane in Figure \ref{fig:Sup fig 4} b. A peak in the transmission appears parallel to the $V_{\mathrm{TG}}$ axis. This feature appears from the part of the device where only the bottom gate does gating to the bilayer graphene. Also, there is some bending of the BLG gap opening feature in the negative displacement sector. We do not fully understand the reason for this bending. 

\subsection{\label{sec:level4.3}S4.3. van-Hove singularity-like feature}
\begin{figure}
    \centering
    \includegraphics[width=\linewidth]{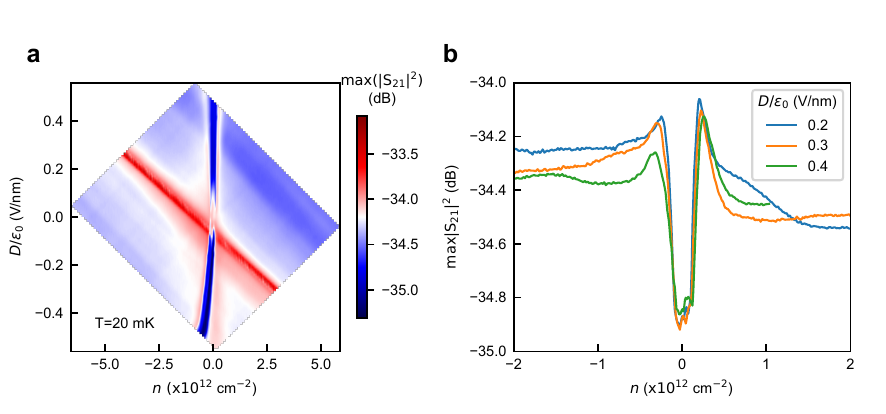}
    \caption {\textbf{Possible signature of van-Hove singularities in BLG.}
    \textbf{a,} For each ($n$, $D/\epsilon_0$) the transmitted signal through the cavity is measured as a function of frequency. The peak value of the amplitude of the transmission coefficient,$|\mathrm{S}_{21}|^2$ is plotted as a function of $n$ and $D/\epsilon_0$.
    \textbf{b,} Shows the line slice of $|\mathrm{S}_{21}|^2$ with $n$ for  $D/\epsilon_0$ = 0.2, 0.3, 0.4 V/nm.
    }
 
    \label{fig:Sup fig 5}
\end{figure}

For each ($n$, $D/\epsilon_0$) point, the transmitted signal through the cavity is measured as a function of frequency. In Figure \ref{fig:Sup fig 5} a, the peak value of the amplitude of the transmission coefficient, $|\mathrm{S}_{21}|^2$ is plotted as a function of $n$ and $D/\epsilon_0$. The line slice of $|\mathrm{S}_{21}|^2$ with n for $D/\epsilon_0$=0.2, 0.3, 0.4 V/nm is shown in Figure \ref{fig:Sup fig 5} b. For all of the $D/\epsilon_0$, we can see a van-Hove singularity-like feature; a peak in the signal next to the gap regions appears. However, the van-Hove singularity-like features are not so prominent in the capacitance ($C_2$) and $d_g$ plots (Figure 4 c and e respectively in the main manuscript). In future works, this could be captured by improving the fitting model and fitting accuracy to extract the capacitance.

As the area of our device is relatively larger $\sim$30 $\mathrm{\mu}$m$^2$, there is an inhomogeneous variation in charge density ($n$). This variation in $n$ may smear out the van Hove singularities at the band edge to some extent\cite{young_electronic_2012}. The observed inhomogeneous variation in
$n$ has been reported in previous studies, contributing to the absence of van Hove singularities in the data \cite{henriksen_measurement_2010}.  

\section{\label{sec:level5}S5. Fitting Procedure}
\begin{figure}[H]
    \centering
    \includegraphics[width=\linewidth]{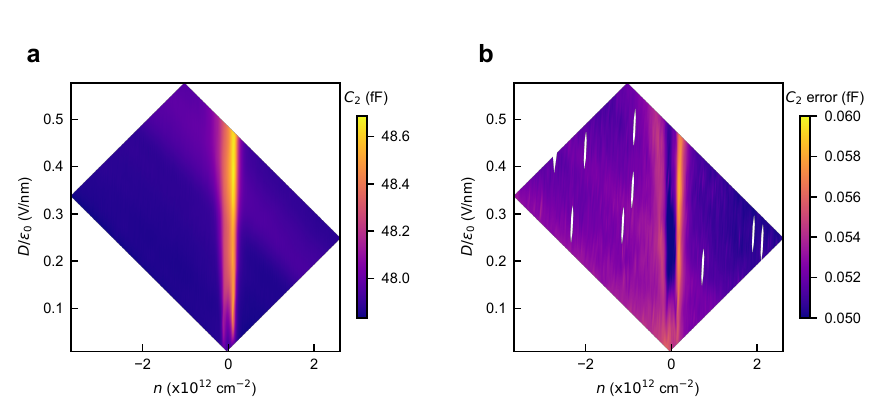}
    \caption {\textbf{Error in the extraction of the capacitance from the fitting.} \textbf{a,} Shows the extracted capacitance $C_2$ from the fitting as a function of $n$ and $D/\epsilon_0$. \textbf{b,} Shows the error in $C_2$ from the fitting.} 
    \label{fig:Sup fig 6}
\end{figure}

First, the frequency line-slice for a particular ($V_{\mathrm{BG}}$, $V_{\mathrm{TG}}$) is fitted with eq 1 in the main text keeping $C_2$, $Q_i$, skewness factor $s$, $b$, and $\omega_0$ as the fitting parameters. From the fitting, we get $\omega_0$=5.285 GHz. Then, for each combination of ($V_{\mathrm{BG}}$, $V_{\mathrm{TG}}$), the frequency line-slice data is subjected to fitting with $C_2$, $Q_i$, skewness factor $s$, and $b$ as the fitting parameters keeping $\omega_0$ constant. The $\omega_0$ is kept constant so that the resonant frequency shifts only due to the capacitive loading of $C_2$, and is not influenced by the change in bare resonant frequency $\omega_0$. Figure \ref{fig:Sup fig 6} a and b show the extracted capacitance $C_2$ and the error in $C_2$ from the fitting respectively. In Figure \ref{fig:Sup fig 6} b there are some white patches, at those points, the error in $C_2$ from the fitting diverges.

To elucidate the impact of capacitive loading on the resonant frequency, we derive the expression for the resonant frequency from eq \ref{Eq:eq1} in the main text under certain assumptions. Substituting the expressions for $A$, $B$, $C$, and $D$ into eq \ref{Eq:eq1} yields:
\begin{equation}\tag{S1}
    \mathrm{S}_{21} = \frac{4 \omega^2 Z_0^2 C_1C_2}{(2j\omega Z_0 C_1+1)(2j\omega Z_0 C_2+1)e^{\gamma l}-e^{-\gamma l}}.
    \label{eq: supp eq1}
\end{equation}

Assuming the $\omega$ is very close to the bare resonant frequency $\omega_0$, the eq \ref{eq: supp eq1} is simplified to
\begin{equation}\tag{S2}
    \mathrm{S}_{21} \approx \frac{\frac{4 \omega^2 Z_0^2 C_1C_2}{\pi}}{\left(\frac{1}{Q_i}-\frac{4 \omega^2 Z_0^2 C_1C_2}{\pi}-\frac{2 \omega Z_0(C_1+C_2)\Delta}{\omega_0}\right) + j\left(\frac{2 \omega Z_0(C_1+C_2)}{\pi}+\frac{2 \Delta}{\omega_0}\right)}.
    \label{eq: supp eq2}
\end{equation}
The resonant frequency, $\omega_r$ can be obtained by setting $\operatorname{Im}(\mathrm{S}_{21})=0$, which gives 
\begin{equation}\tag{S3}
    \omega_r=\omega_0 - \frac{\omega_0 \omega_r Z_0 (C_1 +C_2)}{\pi}.
    \label{eq: supp eq3}
\end{equation}
We can calculate $\omega_r$ by solving eq \ref{eq: supp eq3} iteratively. From eq \ref{eq: supp eq3} it is evident that the resonant frequency $\omega_r$ decreases with an increase in $C_2$. 

\section{\label{sec:level6}S6. Atomic force microscopy of the hBN flakes}
\begin{figure}[H]
    \centering
    \includegraphics[width=\linewidth]{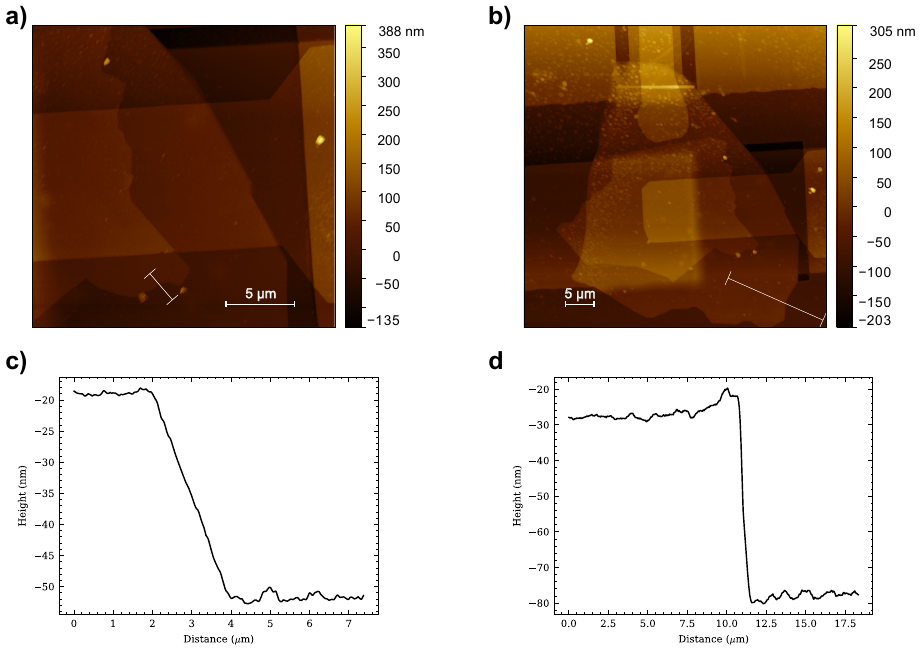}
    \caption {\textbf{AFM data of top and bottom hBNs.}  \textbf{a} and \textbf{b} show the AFM data of the device from which the thickness of the top and bottom hBNs has been extracted respectively. \textbf{c} and \textbf{d} are the line slices along the white line in panel \textbf{a} and \textbf{b} respectively.} 
    \label{fig:Sup fig 7}
\end{figure}

Atomic force microscopy(AFM) of the top and bottom hBN flakes has been performed, which are shown in Figure \ref{fig:Sup fig 7} a and b respectively. Panels c and d show the line slices along the white line in panels a and b respectively. The thicknesses of the top hBN and bottom hBNs are $32.9 \pm  0.6$ nm and $50.1 \pm 1$ nm respectively.
respectively.

\section{\label{sec:level7}S7. Comparison with other works}
In Table \ref{tab:comp_table}, we have compared our technique with other state-of-the-art techniques to measure the capacitances of 2D heterostructures.
\newpage
\begin{longtable}[h] {|p{0.25\linewidth}|p{0.3\linewidth}|p{0.3\linewidth}|p{0.15\linewidth}|}
        \hline
        
        \textbf{Paper}&  \textbf{How the capacitance is measured/extracted}&  \textbf{Accuracy}& \textbf{Frequency range}\\
        \hline

         Electronic compressibility of layer-polarized bilayer graphene, A. F. Young et. al.\cite{young_electronic_2012}& They measured the top gate capacitance (capacitance between the top gate and the graphene) by applying a small AC signal to the top gate and measuring the current through the graphene.& The capacitance measurement has a resolution of sub femtofarad. & Not mentioned\\  
         \hline

         Measurement of the electronic compressibility of bilayer graphene, E. A. Henriksen et. al.\cite{henriksen_measurement_2010} & They measured the penetration field capacitance i.e., they measured the capacitance of the back gate to the top gate, where the electric field lines between the gates penetrate the grounded graphene sheet. They measured it by applying a small AC signal to the back gate and recording the resulting current through the top gate. & They fitted their capacitance data using eq 2 in our main manuscript. They calculated $d_g$ from a tight-binding model and varied the stray capacitance until the best fit was obtained. & 300-1000 Hz \\
         \hline

         Direct measurement of discrete valley and orbital quantum numbers in bilayer graphene, B.M Hunt et. al.\cite{hunt_direct_2017} & They measured both the top gate capacitance (C$_\mathrm{T}$) and bottom gate capacitance (C$_\mathrm{B}$), from which they calculated their symmetric (C$_\mathrm{S}$) and anti-symmetric (C$_\mathrm{A}$) combination as C$_\mathrm{S}$$\equiv$C$_\mathrm{T}$ + C$_\mathrm{B}$, and C$_\mathrm{A}$$\equiv$C$_\mathrm{T}$ - C$_\mathrm{B}$. To measure C$_\mathrm{B(A)}$, two synchronized and nearly equal-magnitude AC signals ($\delta V_\mathrm{EX}$) are applied to the top and bottom gates.  & They have not explicitly mentioned the accuracy or sensitivity of their capacitance measurement. They used a standard capacitor C$_\mathrm{std}$ = 404 fF. The measured C$_\mathrm{S}$ varies between 141.4 fF and 260.2 fF and C$_\mathrm{A}$ varies between -1.6 fF and 1.6 fF. & 67.778 kHz \\
         \hline 

         Spin–orbit-driven band inversion in bilayer graphene by the van der Waals proximity effect, J. O. Island et. al.\cite{island_spinorbit-driven_2019} & They measured C$_\mathrm{P}$ between the top and bottom gates of their BLG-WSe$_2$ heterostructure using a low-temperature capacitance bridge. C$_{\mathrm{P}}$ is measured by applying a fixed AC excitation to the top gate ($\delta \mathrm{V}_{\mathrm{TOP}}$). The phase and amplitude of a second AC excitation with the same frequency are adjusted and applied to a standard reference capacitor (C$_\mathrm{ref}$) on the low-temperature amplifier to balance the capacitance bridge. & They have plotted the C$_{\mathrm{P}}$ in arbitrary units but have not mentioned the accuracy in their capacitance measurement. & 17-33 kHz\\
         \hline

          Half- and quarter-metals in rhombohedral trilayer graphene, H. Zhou et. al.\cite{zhou_half-_2021} & They measured the penetration field capacitance C$_{\mathrm{P}}$ between the top and bottom gates of their device using a low-temperature capacitance bridge by applying a small AC excitation to the top gate. & Measurement accuracy is not mentioned. & 10-55 kHz \\
          \hline

          Isospin magnetism and spin-polarized superconductivity in Bernal bilayer graphene, H. Zhou et. al.\cite{zhou_isospin_2022} & They measured the penetration field capacitance C$_{\mathrm{P}}$ of their bilayer graphene device using a low-temperature capacitance bridge by applying a small AC excitation to the top gate. & Measurement accuracy is not mentioned. & 54.245 KHz \\
          \hline

          This work & We measured the penetration field capacitance C$_{\mathrm{P}}$ of the BLG heterostructure coupled with a superconducting coplanar waveguide cavity. By fitting the transmission data through the cavity using the ABCD transmission matrix model, we extract the value of the capacitance of the heterostructure. & The measured capacitance varies between ~47.8 fF to ~48.7 fF. The error in the capacitance from the fitting varies between 0.05 fF to 0.06 fF. & 1-10 GHz \\
          \hline
    \caption{Comparison table}
    \label{tab:comp_table}
\end{longtable}

\section{\label{sec:level8}S8. Density of states calculation of bilayer graphene}
\begin{figure}[H]
    \centering
    \includegraphics[width=0.7\linewidth]{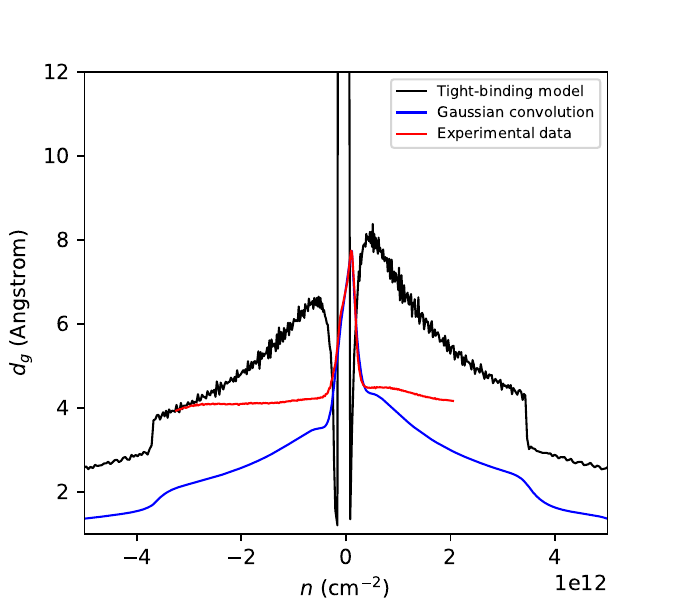}
    \caption {\textbf{Comaparison with theory.} The black curve is the calculated $d_g$ of BLG from the tight-binding model. The $d_g=\frac{\epsilon_0}{e^2}\frac{\partial\mu}{\partial n}$ is obtained from the tight-binding model of BLG\cite{mccann_electronic_2013} in presence of a perpendicular displacement field $D/\epsilon_0$=0.3 V/nm and inter-layer coupling $\gamma_4$=0.14 eV. The blue curve is the convolution of $d_g$ with a Gaussian function with variance $\delta n=1.6\times10^{11}~\mathrm{cm}^{-2}$. The red curve is the experimentally derived $d_g$ at  $D/\epsilon_0$=0.3 V/nm. 
    }
    \label{fig:Sup fig 8}
\end{figure}
To compare our data with theory, we calculated the density of states ($\frac{\partial n}{\partial \mu}$) of BLG from the tight-binding model\cite{mccann_electronic_2013}, from which we derived $d_g$ with the formula $d_g=\frac{\epsilon_0}{e^2}\frac{\partial\mu}{\partial n}$. In Figure \ref{fig:Sup fig 8}, the black curve shows the calculated $d_g$ from the tight-binding model in the presence of a displacement field $D/\epsilon_0$=0.3 V/nm, and inter-layer coupling $\gamma_4$ =0.14 eV. Since our device area is relatively larger $\sim$30 $\mathrm{\mu m}^2$, there is a variation of $n$ across the device. To incorporate this, we have convolved the calculated $d_g$ with a Gaussian with variance, $\delta n=1.6\times10^{11} \mathrm{cm}^{-2}$, which is shown in the blue curve. The Gaussian convolution smoothens the van-Hove singularity features appearing on the band edge of BLG. This matches qualitatively well with the experimental data (shown in the red curve) at low densities, but at higher densities, the matching is increasingly poor, as reported in previous studies\cite{henriksen_measurement_2010}.

\end{document}